\documentclass[12pt]{article}
\usepackage{amsfonts}
\usepackage[mathscr]{eucal}

\newcommand{\be}{\begin{equation}}
\newcommand{\ee}{\end{equation}}
\newcommand{\bea}{\begin{eqnarray}}
\newcommand{\eea}{\end{eqnarray}}

\newcommand{\p}[1]{(\ref{#1})}

\newcommand{\+}{\dagger}

\begin{document}

\begin{titlepage}

\vspace*{1.5cm}

\renewcommand{\thefootnote}{\dag}
\begin{center}

{\LARGE\bf Supersymmetric hyperbolic }
\vspace{0.3cm}

{\LARGE\bf Calogero-Sutherland models }
\vspace{0.3cm}

{\LARGE\bf by gauging}



\vspace{1.5cm}

{\large\bf Sergey~Fedoruk}${}^{\ a}$,\quad {\large\bf Evgeny~Ivanov}${}^{\ a}$,\quad {\large\bf Olaf~Lechtenfeld}${}^{\ b}$
 \vspace{0.5cm}

${}^{\ a}$ {\it Bogoliubov Laboratory of Theoretical Physics, }\\
{\it Joint Institute for Nuclear Research,}\\
{\it 141980 Dubna, Moscow region, Russia} \\
\vspace{0.1cm}

{\tt fedoruk,eivanov@theor.jinr.ru}\\
\vspace{0.5cm}


${}^{\ b}$ {\it Institut f\"ur Theoretische Physik and Riemann Center for Geometry and Physics, }\\
{\it Leibniz Universit\"at Hannover, Appelstra{\ss}e 2, 30167 Hannover, Germany} \\
\vspace{0.1cm}

{\tt lechtenf@itp.uni-hannover.de}\\

\end{center}
\vspace{1cm} \vskip 0.6truecm \nopagebreak

\begin{abstract}
\noindent
Novel $\mathcal{N}{=}\,2$ and $\mathcal{N}{=}\,4$ supersymmetric
extensions of the Calogero-Sutherland hyperbolic systems are
obtained by gauging the ${\rm U}(n)$ isometry of matrix superfield models.
The bosonic core of the $\mathcal{N}{=}\,2$ models is the standard
$A_{n-1}$ Calogero-Sutherland hyperbolic system, whereas the
${\mathcal N}{=}\,4$ model contains additional semi-dynamical spin variables
and is an extension of the U(2) spin Calogero-Sutherland hyperbolic system.
We construct two different versions of the ${\mathcal N}{=}\,4$ model,
with and without the interacting center-of-mass coordinate in the bosonic sector.
\end{abstract}

\vspace{1cm}


\newpage

\end{titlepage}

\setcounter{footnote}{0}
\setcounter{equation}0
\section{Introduction}

Multi-particle Calogero models \cite{C} occupy a distinguished place among the integrable systems.
Most studied  are the rational Calogero (or Calogero-Moser) models respecting one-dimensional
conformal invariance (see \cite{Poly-rev} for a review).
The ${\mathcal N}$-extended supersymmetric versions of such systems are addressed in numerous works
(see, for example, \cite{AP,FR,IKL2,7,nscm,ikl,IvKrN,ikl1} and the review \cite{superc}).
Beyond $\mathcal{N}{=}\,2$ however,
some difficulties in a direct construction of minimal extensions of $n$-particle models
(with ${\mathcal N} n$ fermionic variables) were encountered (see, e.g., \cite{Wyl,GLP-2007,Feig}).
A different approach was put forward in \cite{FIL08} (see also \cite{Fed-2010,superc}),
where ${\mathcal N}{=}\,1,2,4$
supersymmetric extensions of the rational Calogero model were derived by a gauging procedure \cite{DI-06-1,DI-06-2}
applied to matrix superfield systems. Recently, such an approach
was applied for obtaining new superconformal Calogero-Moser systems with deformed
supersymmetry and intrinsic mass parameter \cite{FI,FILS}.
A characteristic feature of the gauging approach is the presence of extra fermionic fields
(as compared to a minimal extension) and, in the ${\mathcal N}{=}\,4$ case,
of bosonic semi-dynamical spinning variables. Within the Hamiltonian formalism,
such a type of matrix system with an extended set of fermionic fields was further utilized
in \cite{KLS-18,KLS-18a,KLS-18b} for deriving general-${\mathcal N}$ supersymmetric rational Calogero models.

Rational Calogero systems constitute only one of the six types of multi-particle integrable systems
classified in \cite{OP,Per}.
In addition to two rational Calogero systems (with and without oscillator potential),
there also exist the Calogero-Sutherland trigonometric system with a potential
proportional to the inverse square of the sine of the coordinate differences,
the Calogero-Sutherland hyperbolic system with a potential proportional to the square
of the hyperbolic sine of the coordinate differences, the Calogero elliptic system with a potential
proportional to the Weierstra\ss{} function of the coordinate differences, as well as a wide class of Toda systems.

To date, supersymmetric generalizations of the Calogero-Sutherland
systems \cite{Su} are understood rather badly. Their
$\mathcal{N}{=}\, 2$ supersymmetric generalizations were constructed
at the component level
\cite{SSuth,BrinkTurW,BorManSas,IoffeNee,DeLaMa}.
For a relation to the representations of diverse superalgebras see~\cite{Serg,SergVes}.
A restricted set of  $\mathcal{N}{=}\,4$ Calogero-Sutherland models with purely quantum potentials
was obtained recently in~\cite{Feig} for $BC_n$, $F_{4}$ and $G_{2}$ root systems.

In this paper we apply the gauging procedure to superfield matrix systems
to obtain the $\mathcal{N}n^2$-type extensions of the hyperbolic $A_{n-1}$ Calogero-Sutherland model,
up to the $\mathcal{N}{=}\,4$ case.
Such a system represents a particular limit of the elliptical Calogero system.
The latter was used for the non-perturbative description of the vacuum sector of $\mathcal{N}{=}\,2, d{=}4$
supersymmetric gauge theories with matter in the Seiberg-Witten setting (see, e.g., \cite{DHoPh,GorMir}).

The plan of the paper is as follows. In Section~2 we show how to obtain the standard bosonic
Calogero-Sutherland hyperbolic system through the matrix-gauging procedure.
In Section~3 we construct a superfield formulation of the $\mathcal{N}{=}\, 2$ supersymmetric
Calogero-Sutherland hyperbolic model. This model is, in some sense,
a direct  $\mathcal{N}{=}\, 2$ superfield generalization of the bosonic matrix model of Section~2.
In contrast to the models of \cite{BrinkTurW,BorManSas,IoffeNee,DeLaMa,SSuth},
this new $\mathcal{N}{=}\, 2$ Calogero-Sutherland hyperbolic model involves $2n^2$ fermions,
like the $\mathcal{N}{=}\, 2$ rational matrix models of \cite{FIL08,superc,DI-06-1,DI-06-2}.
In Section~4 we consider  the $\mathcal{N}{=}\, 4$ superfield model in the $d{=}1$ harmonic superspace formulation.
It essentially includes semi-dynamical U(2) spin variables and provides an $\mathcal{N}{=}\, 4$
supersymmetric generalization of the $\mathrm{U}(2)$-spin Calogero-Sutherland hyperbolic system.
We construct two different versions of the $\mathcal{N}{=}\,4$ model.
The first one is a direct analog of the $\mathcal{N}{=}\,4$ U(2) rational Calogero model of \cite{FIL08},
with an unremovable dependence on the center-of-mass coordinate.
The second one secures a full decoupling of the center of mass in the bosonic sector.

\setcounter{equation}0
\section{Hyperbolic Calogero-Sutherland model by gauging}

Here we consider a modification of the matrix formulation of the $n$-particle rational Calogero model
with $\mathrm{U}(n)$ gauge symmetry given in \cite{FIL08} (see also \cite{Poly-gauge,Gorsky}).
This slight modification will allow us to derive  the hyperbolic Calogero-Sutherland model.
Like in the case of rational models, the basic tool will be an Hermitian matrix $d{=}1$ field.
The use of the Hermitian matrix model for describing
the Calogero-Sutherland hyperbolic system was earlier discussed in~\cite{OP,Per}.\footnote{
For an alternative derivation from free motion on the $\mathrm{GL}(n,\mathbb{C})$ group manifold see~\cite{FeKl}.}
As shown in \cite{Poly-rev}, the analogous construction for the Calogero-Sutherland trigonometric system
requires employing unitary matrices and so is more complicated.\footnote{
An extension of our approach to the trigonometric models will be considered elsewhere.}

The matrix model we will deal with is underlaid by the positive definite Hermitian $n\times n$--matrix field
$$
X(t):=\|X_a{}^b(t)\|\,, \qquad ({X_a{}^b})^* =X_b{}^a\,,\qquad \det X \neq 0\,,
$$
$a,b=1,\ldots ,n$, and the complex $\mathrm{U}(n)$-spinor field
$$
Z(t):=\|Z_a(t)\|\,, \qquad \bar Z^a = ({Z_a})^*\,.
$$
It also involves $n^2$ gauge fields
$$
A(t):=\|A_a{}^b(t)\| \,, \qquad ({A_a{}^b})^* =A_b{}^a\,.
$$
The gauge-invariant action has the following form
\begin{equation}\label{b-Cal}
S =
\int \mathrm{d}t L\,,\qquad L = \frac12\, {\rm Tr}\Big(X^{-1}\nabla\! X \, X^{-1}\nabla\! X \Big) +
\frac{i}{2}\, \Big(\bar Z \nabla\! Z -
\nabla\! \bar Z Z\Big) + c\,{\rm Tr} A   \, ,
\end{equation}
where the covariant derivatives are
\begin{equation}\label{cov-der-b}
\nabla\! X = \dot X +i\, [A, X]\,, \qquad \nabla\! Z = \dot Z + iAZ\,, \qquad \nabla\!
\bar Z
= \dot{\bar Z} -i\bar Z A\,.
\end{equation}
The last term (Fayet-Iliopoulos term) includes only $U(1)$ gauge field, $c$ being a
real constant.

The action (\ref{b-Cal}) is invariant with respect to the local $U(n)$
transformations, $g(\tau )\in U(n)$,
\begin{equation}\label{Un-tran}
X \rightarrow \, g X g^\+ \,, \qquad  Z \rightarrow \, g Z \,, \quad \bar Z
\rightarrow \,
\bar Z g^\+\,,
\qquad
A \rightarrow \, g A g^\+ +i \dot g g^\+\,.
\end{equation}
Using the gauge transformations (\ref{Un-tran}) and
the standard representation $X=URU^\+$ for Hermitian matrix $X$, with $U$ an unitary matrix and $R$ a diagonal matrix,
we can impose a (partial) gauge-fixing
\begin{equation}\label{X-fix}
X_a{}^b =0\,,\qquad a\neq b\,.
\end{equation}
In the gauge (\ref{X-fix}) the matrix variable $X$ takes the form
\begin{equation}\label{X-fix-com}
X_a{}^b = x_a \delta_a{}^b \,.
\end{equation}
Then $[X, A]_a{}^b = (x_a - x_b)A_a{}^b$,
and, therefore,
${\rm Tr}[X, A] = 0$, ${\rm Tr}(\dot X [X, A]) = 0$,
${\rm Tr}([X,A][X, A]) =
-\sum_{a,b}(x_a - x_b)^2 A_a{}^b A_b{}^a$.
As a result, the Lagrangian (\ref{b-Cal}) becomes
\begin{equation}\label{b-Cal2}
L =  \frac12\, \sum_{a,b} \,\left[ \frac{\dot x_a \dot x_a}{(x_a)^2}
+ i (\bar Z^a \dot Z_a - \dot{\bar Z}{}^a Z_a) +
\frac{(x_a - x_b)^2 A_a{}^b A_b{}^a}{x_a x_b} - 2\bar Z^a A_a{}^b Z_b  + 2c\,
A_a{}^a\,\right] .
\end{equation}

The third term in (\ref{b-Cal2}) involves only non-diagonal
elements of the
matrix $A$, $A_a{}^b$ with $a\neq b$. Therefore the Lagrangian (\ref{b-Cal2}) possesses a
residual invariance under the gauge abelian $[U(1)]^n$ group with local
parameters $\varphi_a(t)$:
\begin{equation}\label{b-Ab}
Z_a \rightarrow \, \mathrm{e}^{i\varphi_a} Z_a \,, \quad \bar Z^a  \rightarrow \,
\mathrm{e}^{-i\varphi_a}
\bar Z^a\,, \qquad A_a{}^a \rightarrow \, A_a{}^a - \dot \varphi_a \quad (\mbox{no
sum over}\; a)\,.
\end{equation}
Using this residual gauge invariance, we can impose the further gauge condition
\begin{equation}\label{g-Z}
\bar Z^a = Z_a\,.
\end{equation}
In this gauge, the second term in the Lagrangian (\ref{b-Cal2}) vanishes and the action
takes the form
\begin{equation}\label{b-Cal3}
S = \frac12 \int \mathrm{d}t  \sum_{a,b} \Bigg[\,\frac{\dot x_a \dot x_a}{(x_a)^2}
+ \frac{(x_a - x_b)^2 A_a{}^b A_b{}^a}{x_a x_b} - 2 Z_a Z_b A_a{}^b  + 2c\, A_a{}^a\,\Bigg]\,.
\end{equation}

The equation of motion for $A_a{}^b$ amounts to the relations
\begin{eqnarray}\label{eqA}
A_a{}^b &=& \frac{x_a  x_b}{(x_a - x_b)^2}\, Z_a Z_b  \qquad\quad \mbox{for } a\neq b\,,
\\
\label{eqZ}
(Z_a)^{\,2} &=&c \qquad\quad \forall \, a  \qquad (\mbox{no sum over} a)\,.
\end{eqnarray}
The conditions (\ref{eqZ}) imply $c>0$. Inserting (\ref{eqA}) and (\ref{eqZ})
in the action (\ref{b-Cal2}), we obtain the final gauge-fixed action
\begin{equation}\label{st-Cal}
S = \frac12 \int \mathrm{d}t  \Bigg[\,\sum_{a}\frac{\dot x_a \dot x_a}{(x_a)^2} -
\sum_{a\neq b} \frac{x_a x_b \, c^2}{(x_a - x_b)^2}\,\Bigg]\,.
\end{equation}

Introducing the variables $q_a$ as
\begin{equation}\label{x-q}
x_a=\mathrm{e}^{\,q_a} \,,
\end{equation}
we cast (\ref{b-Cal}) in the form
\begin{equation}\label{el-Cal}
S = \frac12 \int \mathrm{d}t  \Bigg[\, \sum_{a}\dot q_a \dot q_a -
\sum_{a\neq b} \frac{c^2}{4\sinh^2 \frac{q_a - q_b}{2}}\,\Bigg]\,,
\end{equation}
which is just the standard action of the hyperbolic Calogero-Sutherland system of the $A_{n-1}$-root type \cite{Su,OP,Per}.

Note that imposing the gauge-invariant condition
\begin{equation}\label{con-X}
\det X = 1
\end{equation}
in the action (\ref{b-Cal}), we will end up with the hyperbolic Calogero-Sutherland system
described by the action (\ref{el-Cal}) without the center-of-mass coordinate, {\it i.e.} with $\sum_a q_a=0$.

In the next sections we will consider supersymmetric generalizations of the bosonic construction presented here. We will limit our study to the most interesting
${\mathcal N}{=}\,2$ and ${\mathcal N}{=}\,4$ Calogero-Sutherland systems.

\setcounter{equation}{0}
\section{${\mathcal N}{=}\,2$ supersymmetric extension }

\subsection{${\mathcal N}{=}\,2$ superfield action and symmetries }

We will use the following basic ${\mathcal N}{=}\,2$ superfields: the hermitian $n\times n$--matrix superfield
$$
\mathscr{X}_a{}^b(t, \theta,\bar\theta)\,, \qquad (\mathscr{X})^\+ =\mathscr{X}\,,
$$
and chiral $\mathrm{U}(n)$-spinor superfield
$$
\mathcal{Z}_a (t_{\!\scriptscriptstyle{L}}, \theta) \,, \qquad \bar \mathcal{Z}^a
(t_{\!\scriptscriptstyle{R}}, \bar\theta) = (\mathcal{Z}_a)^\+\,,\qquad
t_{\!\scriptscriptstyle{L,R}}=t\pm i\theta\bar\theta\,
$$
$a,b=1,\ldots ,n$, satisfying
\begin{equation}\label{N2-chir-Z}
\bar D \mathcal{Z}_a=0 \,, \qquad D \bar \mathcal{Z}^a=0.
\end{equation}
Here, the covariant spinor derivatives are defined as
$$
D = \partial_{\theta} -i\bar\theta\partial_{t}\,, \quad \bar D = -\partial_{\bar\theta}
+i\theta\partial_{t}\,, \quad \{D, \bar D \} = 2i \partial_{t}\,.
$$
Gauge superfields in the ${\mathcal N}{=}\,2$ case are encompassed by $n\times n$ complex ``bridge'' matrix
$$
b_a{}^b(t, \theta,\bar\theta)\,, \qquad \bar b_a{}^b := {(b_b{}^a)^*}\,
\qquad (\bar b := b^\+).
$$

Let us consider the following gauge invariant action
\begin{equation}\label{N2-Cal-v}
S^{N=2} = \frac12 \int \mathrm{d}t \mathrm{d}^2\theta \,\Bigg[\, {\rm Tr} \left( \mathscr{X}^{-1}\bar\mathscr{D}
\mathscr{X}
\,  \mathscr{X}^{-1}\mathscr{D} \mathscr{X}\, \right) - \bar \mathcal{Z}\, \mathrm{e}^{2V}\!
\mathcal{Z} + 2 c\,{\rm Tr} V \,\Bigg] \,.
\end{equation}
Here the covariant derivatives of the superfield $\mathscr{X}$ are defined by
\begin{equation}\label{cov-der-v}
\mathscr{D} \mathscr{X} =  D \mathscr{X} -i [ \mathscr{A} , \mathscr{X}] \,, \qquad
\bar\mathscr{D} \mathscr{X} = \bar D \mathscr{X} -i [ \bar\mathscr{A} , \mathscr{X}]
\end{equation}
where the gauge potentials are expressed through the bridges as
\begin{equation}\label{pot}
\mathscr{A} =  i \, \mathrm{e}^{i\bar b} (D \mathrm{e}^{-i\bar b}) \,, \qquad D\mathscr{A} = i \,
\mathscr{A}\mathscr{A}\,,
\end{equation}
\begin{equation}\label{b-pot}
\bar\mathscr{A} = i \, \mathrm{e}^{ib} (\bar D \mathrm{e}^{-ib})\,, \qquad \bar D\bar\mathscr{A} = i
\, \bar\mathscr{A}\bar\mathscr{A}\,.
\end{equation}
The gauge prepotential $V$ is defined as
\begin{equation}\label{prepot}
\mathrm{e}^{2V} = \mathrm{e}^{-i\bar b} \, \mathrm{e}^{ib} \,.
\end{equation}

The action~(\ref{N2-Cal-v}) is invariant with respect to the two
types of the local $\mathrm{U}(n)$ transformations:
\begin{itemize}
   \item $\tau$--transformations with the hermitian $n\times n$--matrix parameter
$$
\tau_a{}^b(t, \theta,\bar\theta) \in u(n) \,, \qquad (\tau)^\+ =\tau\,;
$$
   \item $\lambda$--transformations with $n^2$ complex gauge parameters, $\lambda=
\|\lambda_a{}^b\|$,
   $a,b=1,\ldots ,n$, which are (anti)chiral superfields
$$
\lambda (t_{\!\scriptscriptstyle{L}}, \theta) \in u(n) \,, \qquad \bar\lambda
(t_{\!\scriptscriptstyle{R}}, \theta) = (\lambda)^\+  \in u(n) \,.
$$
\end{itemize}
The gauge transformations leaving the
action~(\ref{N2-Cal-v}) invariant are realized as:
\begin{equation}\label{tran-b}
\mathrm{e}^{ib^\prime} = \mathrm{e}^{i\tau} \, \mathrm{e}^{ib} \mathrm{e}^{-i\lambda}\,, \qquad
\mathrm{e}^{i\bar b^\prime} = \mathrm{e}^{i\tau} \, \mathrm{e}^{i\bar b} \mathrm{e}^{-i\bar\lambda}\,,
\end{equation}
\begin{equation}\label{tran-X}
\mathscr{X}^{\,\prime} =  \mathrm{e}^{i\tau}\, \mathscr{X}\, \mathrm{e}^{-i\tau} \,,
\qquad
\mathcal{Z}^{\prime} =  \mathrm{e}^{i\lambda} \mathcal{Z} \,, \quad
\bar \mathcal{Z}^{\prime} =  \bar \mathcal{Z}\, \mathrm{e}^{-i\bar\lambda}\,.
\end{equation}
The potentials are transformed by the $\tau$--group
\begin{equation}\label{tran-A}
\mathscr{A}^{\,\prime} =  \mathrm{e}^{i\tau}\, \mathscr{A}\, \mathrm{e}^{-i\tau} + i\, e^{i\tau} (D
e^{-i\tau})\,,
\qquad \bar\mathscr{A}^{\,\prime} =  \mathrm{e}^{i\tau}\, \bar\mathscr{A}\, \mathrm{e}^{-i\tau} + i\,
e^{i\tau} (\bar D e^{-i\tau})
\end{equation}
whereas the prepotential $V$ undergoes $\lambda$--transformations
\begin{equation}\label{tran-V}
\mathrm{e}^{2V^\prime} = \mathrm{e}^{i\bar\lambda} \, \mathrm{e}^{2V} \mathrm{e}^{-i\lambda}\,.
\end{equation}
Note that the second term in the action~(\ref{N2-Cal-v}) can be
rewritten as
\begin{equation}\label{2f-ac}
\bar \mathcal{Z}\, \mathrm{e}^{2V}\!\mathcal{Z}=
\bar\Phi\,\Phi\, ,\qquad\qquad \Phi\equiv  \mathrm{e}^{ib}\mathcal{Z}\,, \qquad \bar\Phi
\equiv \bar \mathcal{Z} \mathrm{e}^{-i\bar b}\,,
\end{equation}
where $\Phi$, $\bar\Phi$ are subject to the $\tau$--transformations only,
\begin{equation}\label{tran-phi}
\Phi^{\,\prime} =  \mathrm{e}^{i\tau}\, \Phi\,,
\qquad \bar\Phi^{\,\prime} =  \Phi\, \mathrm{e}^{-i\tau} \,.
\end{equation}

\subsection{Prepotential formulation and Wess-Zumino gauge }

It will be convenient to pass to the new field variables on which only $\lambda$ transformations act (``$\lambda$ frame'').
One defines new Hermitian $n\times n$ matrix superfield
\begin{equation}\label{X-tX}
\mathcal{X} = \mathrm{e}^{-ib} \,\mathscr{X}\, \mathrm{e}^{i\bar b}\,,
\end{equation}
in terms of which the action (\ref{N2-Cal-v}) is rewritten as
\begin{equation}\label{N2-Cal}
S^{N=2} = \frac12 \int \mathrm{d}t \mathrm{d}^2\theta \,\Bigg[\, {\rm Tr} \left( \mathcal{X}^{-1}\bar\mathcal{D}
\mathcal{X}
\,\mathcal{X}^{-1} \mathcal{D} \mathcal{X} \right) - \bar \mathcal{Z}\, \mathrm{e}^{2V}\!
\mathcal{Z} +2c\,{\rm Tr} V \,\Bigg] \,,
\end{equation}
where the covariant derivatives of the superfield $\mathcal{X}$ are defined by
\begin{equation}\label{cov-der-s2}
\mathcal{D} \mathcal{X} =  D \mathcal{X} + \mathrm{e}^{-2V} (D \mathrm{e}^{2V}) \, \mathcal{X} \,, \qquad
\bar\mathcal{D} \mathcal{X} = \bar D \mathcal{X} - \mathcal{X} \, \mathrm{e}^{2V} (\bar D
\mathrm{e}^{-2V})\,.
\end{equation}
The superfields present in the action (\ref{N2-Cal}) undergo only chiral $\lambda$-transformations
\begin{equation}\label{tran-pre}
\mathcal{X}^{\,\prime} =  \mathrm{e}^{i\lambda}\, \mathcal{X}\, \mathrm{e}^{-i\bar\lambda} \,,
\qquad
\mathcal{Z}^{\prime} =  \mathrm{e}^{i\lambda} \mathcal{Z} \,, \quad
\bar \mathcal{Z}^{\prime} =  \bar \mathcal{Z}\, \mathrm{e}^{-i\bar\lambda}\,,
\qquad
\mathrm{e}^{2V^\prime} = \mathrm{e}^{i\bar\lambda} \, \mathrm{e}^{2V} \mathrm{e}^{-i\lambda}\,.
\end{equation}

The component contents of the involved superfields are as follows
\begin{equation}\label{com-V}
V = v + \theta \Phi - \bar\theta \bar\Phi + \theta\bar\theta A \,,
\end{equation}
\begin{equation}\label{com-X}
\mathcal{X} = X + \theta \Psi - \bar\theta \bar\Psi + \theta\bar\theta Y \,,
\qquad
\mathcal{Z} = Z + 2i\theta \Upsilon - i\theta\bar\theta \dot Z \,, \quad
\bar\mathcal{Z} =
\bar Z + 2i\bar\theta \bar\Upsilon + i\theta\bar\theta \dot{\bar Z}\,,
\end{equation}
where $\Psi_a{}^b$, $\bar\Psi_a{}^b=({\Psi_b{}^a})^*$
($\bar\Psi = \Psi^\+$), $\Phi_a{}^b$,
$\bar\Phi_a{}^b=({\Phi_b{}^a})^*$ ($\bar\Phi = \Phi^\+$) and
$\Upsilon_a$, $\bar\Upsilon^a=({\Upsilon_a})^*$ are fermionic
fields.

In order to pass to the component form of the action~(\ref{N2-Cal}), we are as usual led to choose Wess-Zumino gauge
\begin{equation}\label{WZ-2a}
V (t, \theta,\bar\theta) = \theta\bar\theta A (t)\,,
\end{equation}
in which
$$
\mathrm{e}^{2V} = 1+ 2\theta\bar\theta A\,, \qquad \mathrm{e}^{-2V} D \mathrm{e}^{2V} = 2D V= 2\bar\theta A\,,
\qquad
\mathrm{e}^{2V} \bar D \mathrm{e}^{-2V} = -2\bar D V = -2\theta A
$$
and
$$
\mathcal{D} \mathcal{X} =\Psi + \bar\theta( Y -i\dot X +2AX) - \theta\bar\theta (2A\Psi
-i\dot\Psi )\,,\quad \bar\mathcal{D} \mathcal{X} = \bar\Psi + \theta( Y +i\dot X
+2XA) -
\theta\bar\theta (2\bar\Psi A +i\dot{\bar\Psi} )\,.
$$
After substituting these Wess-Zumino-gauge  expressions into the action (\ref{N2-Cal}), integrating there over Grassmann coordinates
($\int \mathrm{d}^2\theta\, (\theta\bar\theta) =1 $)
and eliminating auxiliary component fields by their equations of motion,
$$
Y=-\left(AX+XA+\frac12\,\bar\Psi X^{-1}\Psi -\frac12\,\Psi X^{-1}\bar\Psi \right),\qquad \Upsilon=0\,,
$$
the superfield action~(\ref{N2-Cal}) yields the on-shell component action
\begin{equation}\label{WZ-22}
S^{N=2} = \int \mathrm{d}t L^{N=2}\,,
\end{equation}
\begin{eqnarray}
L^{N=2}  &  = & \frac12\,{\rm Tr}\Big( \,
X^{-1}\nabla\! X \,X^{-1}\nabla\! X\Big)  + \frac{i}{2}\, \Big(\bar Z \nabla\! Z - \nabla\! \bar Z Z\Big) + c\,{\rm Tr} A
\nonumber\\ [5pt]
&&
+ \, \frac{i}{2}\,{\rm Tr} \Big( X^{-1}\bar\Psi X^{-1}\nabla \Psi - X^{-1}\nabla \bar\Psi X^{-1}\Psi \Big)
\label{N2Cal-com}\\  [5pt]
&&
- \, \frac{1}{4}\,{\rm Tr} \Big( X^{-1}\bar\Psi X^{-1}\bar\Psi X^{-1}\Psi X^{-1}\Psi \Big)\,.
\nonumber
\end{eqnarray}
Here, $\nabla\! Z$ and $\nabla\! \bar Z$ are defined as in~(\ref{cov-der-b}) and
\begin{equation}\label{cov-der-Psi}
\nabla \Psi = \dot \Psi +i\, [A,\Psi]\,, \qquad \nabla \bar\Psi = \dot {\bar\Psi} +i\, [A,\bar\Psi]\,.
\end{equation}

We observe that the bosonic part of the model (\ref{WZ-22}) (first line of (\ref{N2Cal-com})) is exactly the action of
the hyperbolic Calogero system (\ref{b-Cal}). Hence we have constructed the sought  ${\mathcal N}{=}\,2$ supersymmetric extension of
hyperbolic Calogero-Sutherland system.

It is interesting that, as opposed to the ${\mathcal N}{=}\,2$ supersymmetrization of the rational Calogero model~\cite{FIL08},
${\mathcal N}{=}\,2$ superextension of the hyperbolic Calogero-Sutherland model (\ref{WZ-22}) involves a term quartic in the
fermionic fields (the last line in (\ref{N2Cal-com})).

The action~(\ref{N2Cal-com}) is invariant with respect to the local
$U(n)$ transformations,
$g(\tau )\in \mathrm{U}(n)$, which act on the bosonic fields according to (\ref{Un-tran}) and on the fermionic fields as
\begin{equation}\label{Un-tran-Psi}
\Psi \rightarrow \, g \Psi g^\+ \,, \qquad  \bar\Psi \rightarrow \, g \bar\Psi g^\+\,.
\end{equation}
We can fix this residual gauge freedom by choosing the gauge (\ref{X-fix}), (\ref{X-fix-com}), in which the
matrix $X$ is diagonal. Then,  by analogy with the action (\ref{b-Cal2}), the action obtained
possesses the invariance under gauge Abelian $[U(1)]^n$ symmetry realized by the local
transformations (\ref{b-Ab}) and
\begin{equation}\label{Psi-Ab}
\Psi_a{}^b \rightarrow \, \mathrm{e}^{i\varphi_a} \bar\Psi_a{}^b \mathrm{e}^{-i\varphi_b}\,, \qquad
\bar\Psi_a{}^b \rightarrow \, \mathrm{e}^{i\varphi_a} \bar\Psi_a{}^b  \mathrm{e}^{-i\varphi_b}\,.
\end{equation}
Finally, we can fix the gauge (\ref{g-Z})
\begin{equation}\label{N2-fZ}
\bar Z^a = Z_a \,,
\end{equation}
after which the variables $Z_a$ can be eliminated by their equations of motion.

We would like to point out that the ${\mathcal N}{=}\,0$ and ${\mathcal N}{=}\,2$ gauge constructions above differ
from those relevant to the rational (super)Calogero models merely by the choice of the initial structure of the kinetic term
for the matrix field $X$ and its superfield counterpart ${\mathcal X}$. While in the rational case they are free \cite{FIL08,superc},
in the hyperbolic case they are of sigma-model type. The ${\mathcal N}{=}\,1$ hyperbolic model shares this feature and so can be easily
constructed following the same steps as in the ${\mathcal N}{=}\,1$ rational case (described in detail in \cite{superc}),
with the  properly modified kinetic term for the matrix Hermitian ${\mathcal N}{=}\,1$ superfield. The  ${\mathcal N}{=}\,1$ component action can
be recovered from the action \p{WZ-22}, \p{N2Cal-com} by performing there the reduction $\Psi = \bar\Psi$. Note that
the last quartic  term in \p{N2Cal-com} vanishes upon such a reduction.

\setcounter{equation}{0}
\section{$\mathcal{N}{=}\,4$ supersymmetric extensions}

\subsection{Generalities}

We shall construct the hyperbolic model with $\mathcal{N}{=}\,4$ supersymmetry by analogy with the previously considered
${\mathcal N}{=}\,0$ and ${\mathcal N}{=}\,2$ cases, as well as with the $\mathcal{N}{=}\,4$ superextension  of the rational Calogero model.
The relevant action for the basic matrix superfield proves to be a non-trivial generalization of that
in the rational $\mathcal{N}{=}\,4$ model \cite{FIL08,FIL09,FIL10,superc}.

We shall gauge an action of $n^2$ superfields which form Hermitian $n\times n$ matrix $\mathscr{X}$
and describe off-shell {\bf (1,4,3)} multiplets. The superfield $\mathscr{X}$ is subject to the kinematic constrains
\begin{equation}  \label{cons-4N-fr}
D^iD_i \,\mathscr{X}=0\,,\qquad \bar D_i\bar D^i \,\mathscr{X}=0\,,\qquad
[D^i,\bar D_i]\, \mathscr{X}=0\,,
\end{equation}
where
$$
D^i=\frac{\partial}{\partial\theta_i}-i\bar\theta^i \partial_t\,,\qquad
\bar D_i=\frac{\partial}{\partial\bar\theta^i}-i \theta_i \partial_t\,.
$$
These constraints are solved by
\begin{equation}  \label{sing-X0-WZ}
\mathscr{X}(t,\theta_i,\bar\theta^i)= X + \theta_i\Psi^i +
\bar\Psi_i\bar\theta^i +
i\theta^i\bar\theta^k N_{ik}-{\frac{i}{2}}(\theta)^2\dot{\Psi}_i\bar\theta^i
-{\frac{i}{2}}(\bar\theta)^2\theta_i\dot{\bar\Psi}{}^i +
{\frac{1}{4}}(\theta)^2(\bar\theta)^2 \ddot{X}\,,
\end{equation}
where $(\theta)^2:=\theta_k \theta^k$, $(\bar\theta)^2:=\bar\theta^k \bar\theta_k$ and
$X^\dag = X$, $({\Psi^i})^\dagger=\bar\Psi_i$, $({N^{ik}})^\dagger=N_{ik}=N_{(ik)}$
or, using a more detailed notation,
$$
(X_a{}^b)^*=X_b{}^a\,,\qquad ({\Psi^i}_a{}^b)^*=\bar\Psi_i{}_b{}^a\,,\qquad
({N^{ik}}_a{}^b)^*=N_{ik}{}_b{}^a\,.
$$

The Wess-Zumino term (a counterpart of $\mathcal{Z}$--term in ${\mathcal N}{=}\,2$ case) will be described
by $n$ constrained superfields forming ${\rm SU}(n)$ spinor and describing $n$ {\bf (4,4,0)} multiplets.
In the $d{=}1$ harmonic superspace (HSS) approach \cite{IL} they are represented by $n$ commuting analytic superfields.
Before going further, it is instructive to recall the necessary elements of the $d{=}1$ HSS formalism.

The ${\cal N}{=}\,4$, $d{=}1$ HSS (in the ``central'' basis) is parametrized by the coordinates
$$
(t,\theta^\pm, \bar\theta^\pm, u_i^\pm)\,, \qquad \theta^\pm=\theta^i u_i^\pm
\,, \qquad \bar\theta^\pm=\bar\theta^i u_i^\pm\,,\qquad u^{+i}u_i^-=1\,.
$$
Harmonic analytic superspace is formed by the coordinate set
$$
(\zeta,u)=(t_A,\theta^+, \bar\theta^+, u_i^\pm)\,, \qquad t_A=t+i(\theta^+
\bar\theta^- +\theta^-\bar\theta^+)\,.
$$
The analytic basis in ${\cal N}{=}\,4$, $d{=}1$ HSS is given by the coordinate set
$$
[(\zeta, u), \theta^-, \bar\theta^-].
$$

The invariant integration measures are defined as
$$
\mu_H =\mathrm{d}u\mathrm{d}t\mathrm{d}^4\theta =\mu^{(-2)}_A(D^+\bar D^+)\,,
\qquad
\mu^{(-2)}_A=\mathrm{d}u\mathrm{d}\zeta^{(-2)} =dudt_A \mathrm{d}\theta^+\mathrm{d}\bar\theta^+ =\mathrm{d}u\mathrm{d}t_A(D^-\bar D^-)\,.
$$
The covariant derivatives in the analytic basis are given by the expressions:
\begin{equation}\label{D-ferm}
D^+ =\frac{\partial}{\partial\theta^-}\,,\quad
\bar D^+ =-\frac{\partial}{\partial\bar\theta^-}\,,\qquad
D^- =-\frac{\partial}{\partial\theta^+}-2i\bar\theta^-\partial_{t_A}\,,\quad
\bar D^- =\frac{\partial}{\partial\bar\theta^+}-
2i\theta^-\partial_{t_A}\,,
\end{equation}
\begin{equation}\label{D-harm+}
D^{\pm\pm} =\partial^{\pm\pm} +2i\theta^\pm\bar\theta^\pm\partial_{t_A} +
\theta^\pm\frac{\partial}{\partial\theta^\mp} +\bar\theta^\pm\frac{\partial}{\partial\bar\theta^\mp}\,.
\end{equation}

Below we consider two ${\cal N}{=}\,4$ supersymmetric models
which yield ${\cal N}{=}\,4$ generalizations of the hyperbolic Calogero-Sutherland system.
In first case this ${\cal N}{=}\,4$ generalization contains additional interactions in the bosonic sector including the center of mass.
In second case the ${\cal N}{=}\,4$ model produces only the $\mathrm{U}(2)$-spin hyperbolic Calogero-Sutherland system,
with the center-of-mass coordinate fully decoupled.

\subsection{System \textrm{I} }

\subsubsection{Action}

Now we are prepared for describing the basic ingredients of our ${\mathcal N}=4$ hyperbolic Calogero model.

The ${\cal N}{=}\,4$ supersymmetric and ${\rm U}(n)$ gauge invariant  action is defined as
\begin{equation}\label{4N-gau}
S^{N=4} =S_{\mathscr{X}} + S_{WZ} + S_{FI}\,.
\end{equation}

The first term in~(\ref{4N-gau}) is
\begin{equation}\label{4N-X}
S_{\mathscr{X}} ={\frac{a}{2}}\int \mu_H \,{\rm Tr}\, \Big( \ln\mathscr{X}\, \Big)\,,
\end{equation}
where $\mathscr{X}=\|\mathscr{X}_a{}^b\|$ is subjected to the constraints
\begin{equation}  \label{cons-X-g-V}
\mathscr{D}^{++} \,\mathscr{X}=0\,,
\end{equation}
\begin{equation}  \label{cons-X-g}
\mathscr{D}^{+}\mathscr{D}^{-} \,\mathscr{X}=0\,,\qquad
 \bar\mathscr{D}^{+}\bar\mathscr{D}^{-}\, \mathscr{X}=0\,,\qquad
 (\mathscr{D}^{+}\bar\mathscr{D}^{-} +\bar\mathscr{D}^{+}\mathscr{D}^{-})\, \mathscr{X}=0\,,
\end{equation}
which are none other than the gauge-covariant version of the original constraints \p{cons-4N-fr}. The constraint (\ref{cons-X-g-V}) contains the gauge-covariantized
harmonic derivative
\begin{equation}  \label{def-D++}
\mathscr{D}^{++} = D^{++} + i\,V^{++}\,,
\end{equation}
where the gauge connection $V^{++}(\zeta,u)$ is an analytic unconstrained matrix superfield.
The gauge connections of all other involved covariant derivatives are properly expressed through $V^{++}(\zeta,u)$.
In the analytic basis (and the analytic gauge frame) the spinor derivatives $\mathscr{D}^{+}$ and $\bar\mathscr{D}^{+}$ remain ``short'', $\mathscr{D}^{+} = D^+\,, \;
\bar\mathscr{D}^{+} = \bar{D}^{+}\,$.
Note that $\mathscr{X}$ is in adjoined representation of U($n$) and, therefore,
$$
\mathscr{D}^{++} \,\mathscr{X} = D^{++} \,\mathscr{X} + i\,[V^{++} ,\mathscr{X}]\,,
$$
etc.

The last term in~(\ref{4N-gau}) is FI term
\begin{equation}\label{4N-FI}
S_{FI} =-{\frac{ic}{2}}\,\int \mu^{(-2)}_A \,{\rm Tr} \left(V^{++}
\, \right)\, .
\end{equation}

The second term in~(\ref{4N-gau}) is Wess-Zumino term. It has the following explicit form
\begin{equation}\label{4N-VZ}
S_{WZ} = {\frac{b}{2}}\,\int \mu^{(-2)}_A \, \mathcal{V}_0\, \widetilde{\,\Phi}{}^+\,\Phi^+\, .
\end{equation}
Here, $\Phi^+=(\Phi^+_a), a =1, \ldots, n\,,$ and its conjugate are commuting analytic superfields satisfying the constraints
\begin{equation}  \label{cons-Ph-g}
\mathscr{D}^{++} \,\Phi^+=0\,,
\qquad\qquad
\mathscr{D}^{+} \,\Phi^+=0\,,\qquad
 \bar\mathscr{D}^{+}\, \Phi^+=0\,.
\end{equation}

The real superfield $\mathcal{V}_0$ appearing in \p{4N-VZ} is analytic,
\begin{equation}  \label{cons-V0-g}
\mathcal{V}_0=\mathcal{V}_0(\zeta,u)\,,
\qquad\qquad
{D}^{+} \,\mathcal{V}_0=0\,,\qquad
 \bar{D}^{+}\, \mathcal{V}_0=0\,,
\end{equation}
and it is a prepotential solving the constraints for the singlet part of $\mathscr{X}$ defined as:
\begin{equation}  \label{sing-X0}
\mathscr{X}_0 := {\rm Tr} \left( \mathscr{X} \right)\,.
\end{equation}
For the superfield~(\ref{sing-X0}) the constraints~(\ref{cons-X-g})
do not contain any connections and so take the form~(\ref{cons-4N-fr}):
\begin{equation}  \label{cons-X0}
D^i D_i \,\mathscr{X}_0=0\,,\qquad \bar D_i \bar D^i \,\mathscr{X}_0=0\,,\qquad
 [D^i,\bar D_i]\, \mathscr{X}_0=0\,.
\end{equation}
These constraints are solved in terms of $\mathcal{V}_0$~(\ref{cons-V0-g})
through the integral transform \cite{DI-06-2}
\begin{equation}  \label{X0-V0}
\mathscr{X}_0(t,\theta_i,\bar\theta^i)=\int \mathrm{d}u \,\mathcal{V}_0 \left(t_A,
\theta^+, \bar\theta^+, u^\pm \right)
\Big|_{\theta^\pm=\theta^i u^\pm_i,\,\,\, \bar\theta^\pm=\bar\theta^i u^\pm_i}\,.
\end{equation}
In the actions \p{4N-X}, \p{4N-VZ},  $a$ and $b$ are some non-vanishing coupling constants.

The action~(\ref{4N-gau}) is invariant with respect
to the local U($n$) transformations:
\begin{equation}\label{tran4-X}
\mathscr{X}^{\,\prime} =  \mathrm{e}^{i\lambda}\, \mathscr{X}\, \mathrm{e}^{-i\lambda} \,,
\end{equation}
\begin{equation}\label{tran4-Phi}
\Phi^+{}^{\prime} =  \mathrm{e}^{i\lambda} \Phi^+ \,, \qquad
\widetilde{\,\Phi}{}^+{}^{\prime} =  \widetilde{\,\Phi}{}^+\, \mathrm{e}^{-i\lambda}\,,
\end{equation}
\begin{equation}\label{tran4-V}
V^{++}{}^{\,\prime} =  \mathrm{e}^{i\lambda}\, V^{++}\, \mathrm{e}^{-i\lambda}
+ i\, \mathrm{e}^{i\lambda} (D^{++} \mathrm{e}^{-i\lambda})\,,
\end{equation}
\begin{equation}\label{tran4-X0}
\mathscr{X}_0^{\,\prime} =  \mathscr{X}_0 \,, \qquad
\mathcal{V}_0^{\,\prime} =  \mathcal{V}_0\,,
\end{equation}
where
$
\lambda_a{}^b(\zeta, u^\pm) \in u(n)
$
is the ``hermitian'' $n\times n$--matrix parameter, $\widetilde{\lambda} =\lambda$.
Just due to the choice of such an analytic gauge group, the derivatives $\mathscr{D}^{+}$, $\bar\mathscr{D}^{+}$
in the equations (\ref{cons-X-g}) and (\ref{cons-Ph-g}) contain no gauge connections
\begin{equation}  \label{D-sh}
\mathscr{D}^{+} ={D}^{+}\,,\qquad
 \bar\mathscr{D}^{+}=\bar{D}^{+}\,,
\end{equation}
the feature already  mentioned earlier.

\subsubsection{Wess-Zumino gauge}

Using the analytic gauge freedom~(\ref{tran4-V}), we can choose Wess-Zumino gauge for $V^{++}$:
\begin{equation}  \label{WZ-4N}
V^{++} =2i\,\theta^{+}
 \bar\theta^{+}A(t_A)\,,\qquad \mathscr{D}^{++}= D^{++} -2\,\theta^{+}
 \bar\theta^{+}\,A\,.
\end{equation}
{}From the relation $[\mathscr{D}^{++},\mathscr{D}^{--}]={D}^{0}$ we find
\begin{equation}  \label{D--}
\mathscr{D}^{--}= D^{--} -2\,\theta^{-}
 \bar\theta^{-}\,A\,.
\end{equation}
Then,
\begin{equation}  \label{D-}
\mathscr{D}^{-}=[\mathscr{D}^{--},{D}^{+}]= D^{-} +2\,
 \bar\theta^{-}\,A\,, \qquad \bar\mathscr{D}^{-}=[\mathscr{D}^{--},\bar{D}^{+}]= \bar D^{-} +2\,
 \theta^{-}\,A\,.
\end{equation}
So the Wess-Zumino gauge form of the covariant derivatives is as follows
\begin{equation}\label{D-ferm-WZ}
\mathscr{D}^+ =\frac{\partial}{\partial\theta^-}\,,\quad
\bar \mathscr{D}^+ =-\frac{\partial}{\partial\bar\theta^-}\,,\qquad
\mathscr{D}^- =-\frac{\partial}{\partial\theta^+}-2i\bar\theta^-\nabla_{t_A}\,,\quad
\bar\mathscr{D}^- =\frac{\partial}{\partial\bar\theta^+}-
2i\theta^-\nabla_{t_A}\,,
\end{equation}
\begin{equation}\label{D-harm+WZ}
\mathscr{D}^{\pm\pm} =\partial^{\pm\pm} +2i\theta^\pm\bar\theta^\pm\nabla_{t_A} +
\theta^\pm\frac{\partial}{\partial\theta^\mp} +\bar\theta^\pm\frac{\partial}{\partial\bar\theta^\mp}\,,
\end{equation}
where
\begin{equation}\label{cov-t}
\nabla_{t_A} = \partial_{t_A} +i\, A\,.
\end{equation}

The solution of the constraint (\ref{cons-X-g-V}), (\ref{cons-X-g}) in Wess-Zumino gauge (\ref{WZ-4N}) reads
\begin{eqnarray}   \nonumber
\mathscr{X}&=& X + i\,\theta^-\bar\theta^- N^{++} + i\,\theta^+\bar\theta^+ N^{--}
- i \left(\theta^-\bar\theta^+ + \theta^+\bar\theta^-\right) N + \theta^-\bar\theta^-\theta^+\bar\theta^+ D \\ [6pt]
&&
+\ \theta^- \Psi^+ + \bar\theta^- \bar\Psi^+ - \theta^+ \Psi^- - \bar\theta^+ \bar\Psi^- +
\theta^-\bar\theta^- \left(\theta^+ \Omega^+ + \bar\theta^+ \bar\Omega^+\right), \label{X-WZ}
\end{eqnarray}
where,
\begin{equation}\label{N-WZ}
N^{\pm\pm} = N^{ik}u_i^\pm u_k^\pm \,, \quad N = \nabla_{t_A}X + N^{ik}u_i^+ u_k^- \,,
 \qquad D = 2\nabla_{t_A} \nabla_{t_A}X +2 \nabla_{t_A} N^{ik}u_i^+ u_k^- \,,
\end{equation}
\begin{equation}\label{Psi-WZ}
\Psi^{\pm} = \Psi^{i}u_i^\pm  \,,\quad \bar\Psi^{\pm} = \bar\Psi^{i}u_i^\pm \,,\qquad
\Omega^{+} = 2i\nabla_{t_A}\Psi^{i}u_i^+  \,,\quad \bar\Omega^{+} = 2i\nabla_{t_A}\bar\Psi^{i}u_i^+
\end{equation}
and
\begin{equation}\label{fi-X-an}
X(t_A)\,, \qquad N^{ik}= N^{(ik)}(t_A) \,,\qquad \Psi^{i}(t_A) \,, \qquad \bar\Psi^{i}(t_A)
\end{equation}
are usual $d{=}1$ fields having no harmonic dependence.

The constraints~(\ref{cons-Ph-g}) in Wess-Zumino gauge~(\ref{WZ-4N}) are solved by
\begin{equation}  \label{Ph-WZ}
\Phi^+ = Z^+ + \theta^+ \varphi + \bar\theta^+ \phi
- 2i\, \theta^+ \bar\theta^+ Z^- \,,\qquad
\widetilde{\Phi}{}^+ = \widetilde{Z}{}^+ + \theta^+ \bar\phi - \bar\theta^+ \bar\varphi
- 2i\, \theta^+ \bar\theta^+ \widetilde{Z}{}^-\,,
\end{equation}
with
\begin{equation}\label{Z+-WZ}
Z^{+} = Z^{i}u_i^+  \,,\qquad \widetilde{Z}{}^+ = \bar Z_{i}u^{+i} \,,\qquad
\bar Z_{i} = (Z^{i})^* \,,
\end{equation}
\begin{equation}\label{Z--WZ}
Z^{-} = \nabla_{t_A}Z^{i}u_i^-  \,,\qquad \widetilde{Z}{}^- = \nabla_{t_A}\bar Z_{i}u^{-i}\,,
\end{equation}
and
\begin{equation}\label{fi-Ph-an}
Z^{i}(t_A)\,, \qquad \varphi(t_A) \,,\qquad
\phi(t_A)
\end{equation}
being $d{=}1$ fields.

{}Using (\ref{X0-V0}), we can also determine the real analytic superfield $\mathcal{V}_0$.
{}First, from the expression~(\ref{X-WZ}) we find the singlet part $\mathscr{X}_0 = {\rm Tr} (\mathscr{X})$ (in the central basis):
\begin{equation}  \label{sing-X0-WZ0}
\mathscr{X}_0(t,\theta_i,\bar\theta^i)= X_0 + i\theta^i\bar\theta^k N_0{}_{ik} +
{\frac{1}{4}}(\theta)^2(\bar\theta)^2 \ddot{X}_0- \theta^i\Psi_0{}_i
- \bar\theta^i\bar\Psi_0{}_i +{\frac{i}{2}}(\theta)^2\bar\theta^i\dot{\Psi}_0{}_i
+{\frac{i}{2}}(\bar\theta)^2\theta^i\dot{\bar\Psi}_0{}_i\,,
\end{equation}
where
\begin{equation}  \label{def-X0}
X_0 := {\rm Tr} (X)\,,\qquad
N_0^{ik} := {\rm Tr} (N^{ik})\,,\qquad
\Psi_0^i := {\rm Tr} (\Psi^i)\,,\qquad
\bar\Psi_0^i := {\rm Tr} (\bar\Psi^i)\,.
\end{equation}
Then, from~(\ref{X0-V0}) we restore $\mathcal{V}_0$ as
\footnote{Note that $\int \mathrm{d}u \,u^{+i}u^-_k=-\int \mathrm{d}u \,u^{-i}u^+_k=\frac{1}{2}\,\delta^i_k$,
$\int \mathrm{d}u \,u^{+(i_1}u^{+i_2)}u^-_{(k_1} u^-_{k_2)}=
\int \mathrm{d}u \,u^{-(i_1}u^{-i_2)}u^+_{(k_1} u^+_{k_2)}=
-2\int \mathrm{d}u \,u^{+(i_1}u^{-i_2)}u^+_{(k_1} u^-_{k_2)}=
\frac{1}{3}\,\delta^{(i_1}_{(k_1}\delta^{i_2)}_{k_2)}$ and
$(\theta)^2=-2\theta^+\theta^-$, $(\bar\theta)^2=2\bar\theta^+\bar\theta^-$.}
\begin{equation}  \label{V0-WZ}
\mathcal{V}_0 (\zeta, u)
=X_0(t_A) + 3i\,\theta^+ \bar\theta^+ N_0^{ik}(t_A)u^-_i u^-_k -
2\,\theta^+ \Psi_0^{i}(t_A)u^-_i  -
2\,\bar\theta^+ \bar\Psi^{i}_0(t_A)u^-_i\,.
\end{equation}
While finding this expression, we made use of the abelian gauge freedom of the representation (\ref{X0-V0}),
$\mathcal{V}_0{}' = \mathcal{V}_0 + D^{++}\Lambda^{--}$, $\Lambda^{--} = \Lambda^{--}(\zeta, u)\,$.

Inserting the expressions (\ref{X-WZ}), (\ref{Ph-WZ}), (\ref{V0-WZ})
in the action~(\ref{4N-gau}), using the identity
\begin{equation}\label{ident-matr}
\partial \, {\rm Tr}\,\Big(\ln {\mathscr{X}} \Big)\equiv  \partial \ln\,\Big( \det {\mathscr{X}} \Big)=
{\rm Tr}\,\Big({\mathscr{X}}^{-1} \partial {\mathscr{X}} \Big)
\end{equation}
and, finally,  integrating over Grassmann and harmonic coordinates, we obtain
\begin{equation}\label{4N-gau-WZ}
S^{N=4} = S_{\mathscr{X}} + S_{WZ} + S_{FI}\,,
\end{equation}
with
\begin{eqnarray}
\label{4N-X-WZ}
S_{\mathscr{X}} &=&  \frac{a}{2}\, {\rm Tr}\! \int \mathrm{d}t \,\Big( X^{-1}\nabla X X^{-1}\nabla X
+i\,X^{-1}\nabla\bar\Psi_k X^{-1}\Psi^k-i\,X^{-1}\bar\Psi_k X^{-1}\nabla\Psi^k
\\
\nonumber
&&  \quad +\,\frac12\,  X^{-1}N^{ik}X^{-1}N_{ik}
-i\,[X^{-1}\Psi^i, X^{-1}\bar\Psi^k]\, X^{-1}N_{ik}
\\ [6pt]
\nonumber
&&  \quad -\,\frac13\, [X^{-1}\Psi^i, X^{-1}\bar\Psi^k]\, [X^{-1}\Psi_{(i}, X^{-1}\bar\Psi_{k)}]
-\,\frac13\,X^{-1}\Psi^i X^{-1}\Psi^k\, X^{-1}\bar\Psi_{(i} X^{-1}\bar\Psi_{k)}
\\ [6pt]
\nonumber
&&  \quad +\,\frac12\, X^{-1}\Psi^i X^{-1}\Psi_i\, X^{-1}\bar\Psi^{k} X^{-1}\bar\Psi_{k}\,\Big)\, ,
\\ [6pt]
\label{4N-FI-WZ}
S_{FI} &= & c\,\,{\rm Tr} \int \mathrm{d}t \,A\, ,
\\
\label{4N-WZ-WZ}
S_{WZ} &= & \frac{ib}{2}\, \int \mathrm{d}t \, X_0 \left(\nabla \bar Z_k \, Z^k-\bar Z_k \nabla Z^k \right)
- \frac{ib}{2}\, \int \mathrm{d}t \,N_0^{ik}\bar Z_i Z_k
\\
\nonumber
&& +\, \frac{b}{2}\, \int \mathrm{d}t \, \Big[\Psi^k_0 \left(\bar Z_k\phi + \bar\varphi Z_k \right)
+ \bar\Psi^k_0 \left(\bar\phi Z_k- \bar Z_k\varphi\right) - X_0
\left(\bar\phi\phi+ \bar\varphi\varphi\right)\Big].
\end{eqnarray}
In these expressions,
$$
\nabla X =\dot X+i[A,X]\,,\qquad
\nabla \Psi^k =\dot \Psi^k+i[A,\Psi^k]\,,\qquad
\nabla \bar\Psi_k =\dot{\bar\Psi}_k+i[A,\bar\Psi_k]\,,
$$
$$
\nabla Z^k=\dot Z^k + i A  Z^k\,,\qquad
\nabla \bar Z_k=\dot{\bar Z}_k -i \bar Z_k A\,.
$$

The equations of motion for the fields $N^{ik}$, $\phi$,  $\bar\phi$, $\varphi$,  $\bar\varphi$
are algebraic and can be used to eliminate these fields:
\begin{equation}\label{4N-eq-N}
N^{ik} =iX[X^{-1}\Psi^{(i}, X^{-1}\bar\Psi^{k)}]+ \frac{ib}{a} \, (\bar Z^{(i} Z^{k)})X^2\,,
\end{equation}
\begin{equation}\label{4N-eq-phi}
\phi =-\frac{1}{X_0}\, \bar\Psi^{k}_0 Z_{k}\,, \quad
\bar\phi =\frac{1}{X_0}\, \Psi^{k}_0 \bar Z_{k}\,, \quad
\varphi =-\frac{1}{X_0}\, \Psi^{k}_0 Z_{k}\,, \quad
\bar\varphi =-\frac{1}{X_0}\, \bar\Psi^{k}_0 \bar Z_{k}\,.
\end{equation}
Inserting these expressions in~(\ref{4N-gau-WZ}), we obtain
\begin{eqnarray}
S^{N=4} &=&  \frac12\, \int \mathrm{d}t \,\Big[\,{\rm Tr} \left(  aX^{-1}\nabla X X^{-1}\nabla X
+2c \,A \right)+ ib\,X_0 \left(\nabla \bar Z_k \, Z^k - \bar Z_k \nabla Z^k \right)
\nonumber\\ [6pt]
&&  \qquad \qquad +\,{\frac{b^2}{2a}}\,(\bar Z^{(i} Z^{k)})(\bar Z_{i} Z_{k}) \,{\rm Tr}\left(X^2\right)
\Big] \nonumber\\ [6pt]
&&  +\,\frac{a}{2}\, {\rm Tr}\! \int \mathrm{d}t \,\Big(
i\,X^{-1}\nabla\bar\Psi_k X^{-1}\Psi^k-i\,X^{-1}\bar\Psi_k X^{-1}\nabla\Psi^k
\nonumber \\ [6pt]
&&  \qquad -\,\frac16\, [X^{-1}\Psi^i, X^{-1}\bar\Psi^k]\, [X^{-1}\Psi_{(i}, X^{-1}\bar\Psi_{k)}]
-\,\frac13\,X^{-1}\Psi^i X^{-1}\Psi^k\, X^{-1}\bar\Psi_{(i} X^{-1}\bar\Psi_{k)}
\nonumber \\ [6pt]
\nonumber
&&  \qquad +\,\frac12\, X^{-1}\Psi^i X^{-1}\Psi_i\, X^{-1}\bar\Psi^{k} X^{-1}\bar\Psi_{k}\,\Big)
\nonumber\\ [6pt]
&& +\,\frac{b}{2}\, \int \mathrm{d}t \,(\bar Z_{(i} Z_{k)}) \left(
{\rm Tr}\left(X[X^{-1}\Psi^i, X^{-1}\bar\Psi^k]\right) - \frac{2}{X_0}\, \Psi^{i}_0\bar\Psi^{k}_0\right)
.\label{4N-gau-bose-1}
\end{eqnarray}

\subsubsection{Bosonic limit}

Let us consider the bosonic limit of the action~(\ref{4N-gau-bose-1}), {\it i.e.}, the first two lines in it.
For simplicity, we put $a=-b=1\,$.

We also introduce the new fields
\begin{equation}\label{4N-nZ}
Z^\prime{}^i_a =  (X_0)^{1/2}\,Z^i_a\,,
\end{equation}
and, in what follows, omit the primes on the newly defined fields. The bosonic part of the action~(\ref{4N-gau-bose-1})
becomes
\begin{eqnarray}
S_{bose}^{N=4} &=&  \frac12\, \int \mathrm{d}t \,\Big[\,{\rm Tr} \left(  X^{-1}\nabla X X^{-1}\nabla X
+2c \,A \right)+ i \left(\bar Z_k \nabla Z^k - \nabla \bar Z_k \, Z^k \right)
\nonumber\\ [6pt]
&&  \qquad \qquad +\,\frac{(\bar Z^{(i} Z^{k)})(\bar Z_{i} Z_{k}) \,{\rm Tr}\left(X^2\right)}{2(X_0)^2}
\,\Big]\,.\label{4N-gau-bose-2}
\end{eqnarray}

The action (\ref{4N-gau-bose-2}) respects the residual invariance (\ref{tran4-X})--(\ref{tran4-X0})
of Wess-Zumino gauge~(\ref{WZ-4N}):
\begin{equation}\label{r-tran4-X}
X^{\,\prime} =  \mathrm{e}^{i\lambda}\, X\, \mathrm{e}^{-i\lambda} \,,
\end{equation}
\begin{equation}\label{r-tran4-Z}
Z_a^{\prime}{}^{k} =  \mathrm{e}^{i\lambda} Z_a^{k} \,, \qquad
\bar Z_k^{\prime}{}^{a} =  \bar Z_k^{a}\, \mathrm{e}^{-i\lambda}\,,
\end{equation}
\begin{equation}\label{r-tran4-A}
A^{\,\prime} =  \mathrm{e}^{i\lambda}\, A\, \mathrm{e}^{-i\lambda}
+ i\, \mathrm{e}^{i\lambda} (\partial_t \mathrm{e}^{-i\lambda})\,,
\end{equation}
\begin{equation}\label{r-tran4-X0}
X_0^{\,\prime} =  X_0\,,
\end{equation}
where
$
\lambda_a{}^b(t) \in u(n)
$
are the gauge parameters depending only on $t$. Now we can impose the gauge-fixing conditions \p{X-fix}
\begin{equation}\label{4N-X-fix}
X_a{}^b =0\,,\qquad a\neq b.
\end{equation}
As a result, the action~(\ref{4N-gau-bose-2}) acquires the form (the sums over repeating indices $i,k=1,2$ are always assumed)
\begin{eqnarray}\nonumber
S_{bose}^{N=4} &=& \frac12\, \int \mathrm{d}t  \sum_{a,b} \,\Bigg[\, \frac{\dot x_a \dot x_a}{(x_a)^2}
+ i ( \bar Z_k^a \dot Z^k_a - \dot {\bar Z}{}_k^a Z^k_a ) + \frac{(x_a - x_b)^2 A_a{}^b A_b{}^a}{x_a x_b} \\
&& \qquad\qquad\qquad  - 2\bar Z_k^a A_a{}^b Z^k_b  + 2c\,
A_a{}^a+
\frac{\bar Z^a{}^{(i} Z_a^{k)}\,\bar Z^b_{(i} Z_{k)}{}_b\,{\rm Tr}\left(X^2\right)}{2(X_0)^2}\,\Bigg]\,. \label{4N-gau-bose-fix}
\end{eqnarray}
Note that
$
X_0 =\sum_a x_a\,.
$

The equations of motion for $A_a{}^b$ read
\begin{equation}\label{4N-eq-ab}
A_a{}^b = \frac{x_a x_b}{(x_a - x_b)^2}\, Z^i_a \bar Z_i^b  \qquad\qquad\qquad \mbox{for } a\neq b\,,
\end{equation}
\begin{equation}\label{4N-eq-aa}
\qquad\qquad \bar Z_i^a Z^i_a =c \qquad\qquad \forall \, a  \qquad (\mbox{no sum over the index}\;\,
a)\,.
\end{equation}
Substituting this in~(\ref{4N-gau-bose-fix}), we obtain
\begin{eqnarray}\nonumber
S_{bose}^{N=4} &=& \frac12\, \int \mathrm{d}t \Bigg\{ \sum_{a} \,\Big[\, \frac{\dot x_a \dot x_a}{(x_a)^2} +
i (\bar Z_k^a \dot Z^k_a - \dot {\bar Z}{}_k^a Z^k_a) \Big]  \\
&& \qquad - \sum_{a\neq b} \, \frac{x_a x_b\bar Z_i^a Z^i_b\,\bar Z_k^b Z^k_a}{(x_a - x_b)^2}
+
\sum_{a,b} \, \frac{\bar Z^a{}^{(i} Z_a^{k)}\,\bar Z^b_{(i} Z_{k)}{}_b\,{\rm Tr}\left(X^2\right)}{2(X_0)^2}\,\Bigg\}. \label{4N-bose-fix}
\end{eqnarray}
The $d{=}1$ fields $Z^k_a$ are subjected by the constraints~(\ref{4N-eq-aa}) and carry
the residual $[U(1)]^n$ gauge symmetry  with local
parameters $\varphi_a(t)$:
\begin{equation}\label{4N-b-Ab}
Z_a^k \rightarrow \, \mathrm{e}^{i\varphi_a} Z_a^k \,, \qquad \bar Z^a_k  \rightarrow \,
\mathrm{e}^{-i\varphi_a} \bar Z^a_k\,.
\end{equation}

Introducing the variables $q_a$ as in  \p{x-q},
\begin{equation}\label{x-q4}
x_a=\mathrm{e}^{q_a} \,,
\end{equation}
we observe that the action (\ref{4N-bose-fix}) takes the form
\begin{eqnarray}
S_{bose}^{N=4} &=& \frac12\, \int \mathrm{d}t \Bigg\{ \sum_{a} \,\Big[\, \dot q_a \dot q_a +
i (\bar Z_k^a \dot Z^k_a - \dot {\bar Z}{}_k^a Z^k_a) \Big]
- \sum_{a\neq b} \, \frac{(S_a)_i{}^k(S_b)_k{}^i}{4\sinh^2 {\displaystyle\frac{q_a - q_b}{2}}} \nonumber \\
&& + \sum_{a,b} \, \frac{(S_a)^{(ik)}\,(S_b)_{(ik)}\,{\rm Tr}\left(X^2\right)}{2(X_0)^2},\Bigg\} \,,
\label{4Na-bose-fix1}
\end{eqnarray}
where
\begin{equation}
\label{def}
{\rm Tr}\left(X^2\right) = \sum_c \mathrm{e}^{2q_c}\,, \quad X_0 = \sum_c \mathrm{e}^{q_c}
\end{equation}
and
\begin{equation}
\label{def-S}
(S_a)_i{}^k := \bar Z_i^a Z^k_a \,.
\end{equation}
The quantities \p{def-S} generate $n$ copies of $\mathrm{U}(2)$ algebra. As a result, the above action, up to the last term,  describes
the hyperbolic $\mathrm{U}(2)$-spin Calogero-Sutherland system (see, e.g.,  \cite{Poly2001,Poly-rev}).

\subsection{System \textrm{II} }

In the total action (\ref{4N-gau}) the Wess-Zumino term
(\ref{4N-VZ})  involves the interaction of the trace part of the
matrix {\bf (1,4,3)} multiplet with spin {\bf (4,4,0)} multiplets.
Just this Wess-Zumino term was employed in \cite{FIL08} to construct
the $\mathcal{N}{=}\,4$ rational Calogero system. This specific form of
the Wess-Zumino term was chosen exclusively in order to ensure the
$\mathcal{N}{=}\,4$ superconformal invariance of the full action.

However, the $\mathcal{N}{=}\,4$ hyperbolic model we are considering here lacks the superconformal invariance already in the kinetic term (\ref{4N-X}).
So, there are no reasons to require such an invariance  for the Wess-Zumino term as well.
Keeping this in mind, we can alternatively  consider the system with the action
\begin{equation}\label{4N-gau-1}
\tilde S^{N=4} =S_{\mathscr{X}} + S_{WZ}^{\,\prime} + S_{FI}\,,
\end{equation}
with the simplest Wess-Zumino term
\begin{equation}\label{4N-VZ-1}
S_{WZ}^{\,\prime} = {\frac{b}{2}}\,\int \mu^{(-2)}_A \, \widetilde{\,\Phi}{}^+\,\Phi^+\, ,
\end{equation}
and the same $S_{\mathscr{X}}$ and $S_{FI}$ as in (\ref{4N-X}) and
(\ref{4N-FI}).

In the Wess-Zumino gauge (\ref{WZ-4N}) the action  (\ref{4N-VZ-1}) acquires the following component form (cf. (\ref{4N-WZ-WZ}))
\begin{equation}
\label{4N-WZ-WZ-1}
S_{WZ}^{\,\prime} = \frac{b}{2}\, \int \mathrm{d}t \, \Big[i \left(\nabla \bar Z_k \, Z^k-\bar Z_k \nabla Z^k \right)
- \left(\bar\phi\phi+ \bar\varphi\varphi\right)\Big],
\end{equation}
implying the very simple on-shell expressions for the auxiliary fields:
\begin{equation}\label{4N-eq-N-1}
N^{ik} =iX[X^{-1}\Psi^{(i}, X^{-1}\bar\Psi^{k)}]\,, \qquad
\phi =
\bar\phi =
\varphi =
\bar\varphi =0\,,
\end{equation}
in contrast to (\ref{4N-eq-N}), (\ref{4N-eq-phi}).
As a result, the component on-shell form of the action (\ref{4N-gau-1}) is given by
\begin{eqnarray}
\tilde S^{N=4} &=&  \frac12\, \int \mathrm{d}t \,\Big[\,{\rm Tr} \left(  aX^{-1}\nabla X X^{-1}\nabla X
+2c \,A \right)+ ib  \left(\nabla \bar Z_k \, Z^k - \bar Z_k \nabla Z^k \right)
\Big] \nonumber\\ [6pt]
&&  +\,\frac{a}{2}\, {\rm Tr}\! \int dt \,\Big(
i\,X^{-1}\nabla\bar\Psi_k X^{-1}\Psi^k-i\,X^{-1}\bar\Psi_k X^{-1}\nabla\Psi^k
\nonumber \\ [6pt]
&&  \qquad -\,\frac16\, [X^{-1}\Psi^i, X^{-1}\bar\Psi^k]\, [X^{-1}\Psi_{(i}, X^{-1}\bar\Psi_{k)}]
-\,\frac13\,X^{-1}\Psi^i X^{-1}\Psi^k\, X^{-1}\bar\Psi_{(i} X^{-1}\bar\Psi_{k)}
\nonumber \\ [6pt]
&&  \qquad +\,\frac12\, X^{-1}\Psi^i X^{-1}\Psi_i\, X^{-1}\bar\Psi^{k} X^{-1}\bar\Psi_{k}\,\Big)
.\label{4N-gau-bose-1a}
\end{eqnarray}

The bosonic limit of this action (the first line in (\ref{4N-gau-bose-1a})) has already the correct form for the variables $Z^i_a$, so
the rescaling (\ref{4N-nZ}) is unnecessary in this case. After gauge-fixing \p{4N-X-fix} and eliminating the auxiliary fields by the equations \p{4N-eq-ab} at $a=-b=1\,$
we obtain that the bosonic limit of (\ref{4N-gau-bose-1a}) is described by the action
\begin{eqnarray}
\tilde{S}_{bose}^{N=4} = \frac12\, \int \mathrm{d}t \Bigg\{ \sum_{a} \,\Big[\, \dot q_a \dot q_a +
i (\bar Z_k^a \dot Z^k_a - \dot {\bar Z}{}_k^a Z^k_a) \Big]
- \sum_{a\neq b} \, \frac{(S_a)_i{}^k(S_b)_k{}^i}{4\sinh^2 \frac{q_a - q_b}{2}}\Biggr\}\,.
\label{4Na-boseII-fix1}
\end{eqnarray}
Recall that the last term in the previous bosonic action
\p{4N-bose-fix} involves ${\rm Tr}\left(X^2\right)$ and $X_0$. The
latter coordinate (the center-of-mass coordinate) decouples only for
the trivial cases $n=1,2$. In contrast, the action \p{4N-gau-1}
for any $n$  yields in the bosonic sector the pure hyperbolic
$\mathrm{U}(2)$-spin Calogero-Sutherland system \p{4Na-boseII-fix1}
without any additional interaction. The center-of-mass coordinate
is fully detached and described  by the free action in this sector.

Finally, note that in the one-particle case $n{=}1$ in the system \p{4N-gau-1} there comes about a separation of
the dynamic $\mathscr{X}$-sector from the semi-dynamic $\Phi$ sector.
The Lagrangian of the dynamic sector describes a free system, not possessing any potential
(see, e.g., \cite{FIL12}). Non-trivial cases start with the two-particle system $n{=}2$.
In this case the relative-motion sector (no-center-of-mass sector) of the system  \p{4N-gau-1}
with the coordinate $y=(q_1{-}q_2)/2$
describes an $\mathcal{N}{=}\,4$ generalization of a particular (hyperbolic)  P\"{o}schl-Teller system \cite{PoTe}
with additional bosonic spin variables $Z_a^k$. As opposed to the $\mathcal{N}{=}\, 4$ case,
in the $\mathcal{N}{=}\,0$ and $\mathcal{N}{=}\,2$ cases (Sections 2 and 3) the relevant two-particle  models
describe P\"{o}schl-Teller systems without any extra spin variables.\footnote{
For supersymmetric many-particle generalizations of the P\"{o}schl-Teller system see, e.g., \cite{LNY,GNS,HNS,EDN,KKLNS}.}

\setcounter{equation}{0}
\section{Conclusions}

In this paper we have derived new models of multi-particle supersymmetric mechanics,
which are $\mathcal{N}{=}\,2$ and $\mathcal{N}{=}\,4$ supersymmetric generalizations of the
Calogero-Sutherland hyperbolic system of $A_{n-1}$ root type.
The input in our construction was superfield matrix systems with ${\rm U}(n)$ gauge symmetry.
Because of this, the resulting $n$-particle systems feature $\mathcal{N}n^2$ real physical fermions, like the rational
$\mathcal{N}{=}\,2$ and $\mathcal{N}{=}\,4$ Calogero systems constructed in \cite{FIL08,Fed-2010,superc}
by using a similar gauging approach.
In the $\mathcal{N}{=}\,4$ case, the system involves additional semi-dynamical bosonic spin variables
and so describes a $\mathcal{N}{=}\,4$ supersymmetric generalization of the $\mathrm{U}(2)$-spin
Calogero-Sutherland hyperbolic system.

In subsequent works we aim at the explicit form of the supersymmetry
generators in the models constructed here and plan carry out their
quantization. An interesting question is as to whether the classical
and quantum integrability of the bosonic Calogero-Sutherland model
is preserved upon the supersymmetrization considered. Another option
is the generalization of the ${\mathcal N}{=}\,4$ model to deformed
SU(2$|$1) supersymmetry along the lines of \cite{FI,FILS}. The
intrinsic mass parameter (deformation parameter) in this case will
amount to an additional oscillator potential in the corresponding
actions.

Finally, the remaining integrable many-particle Calogero-type systems from the list of \cite{OP,Per} wait to be supersymmetrized.

\bigskip
\section*{Acknowledgements}
The work of S.F.\ and E.I.\ was supported by the Russian Science Foundation, grant no.\,16-12-10306.
We are grateful to Laszlo Feher for useful comments. E.I.\ thanks Armen Nersessian for valuable correspondence.

\end{document}